\documentstyle[prl,aps]{revtex}

\begin{document}
\draft

\twocolumn[\hsize\textwidth\columnwidth\hsize\csname  @twocolumnfalse\endcsname \title{ 
Scaling of the Hall Resistivity in the Solid and Liquid Vortex Phases
 in Twinned Single Crystal YBa$_{2}$Cu$_{3}$O$_{7-\delta }$ }
\author{G. D'Anna,$^{1}$ V. Berseth,$^{1}$ L. Forr\'{o},$^{1}$ A. Erb,$^{2}$ and E. Walker$^{2}$}
\address{
1 Institut de G\'{e}nie Atomique, Ecole Polytechnique F\'{e}d\'{e}rale de
Lausanne, CH-1015 Lausanne, Switzerland
\\
2 D\'{e}partement de Physique de la Mati\`{e}re Condens\'{e}e, Universit\'{e}
de Gen\`{e}ve, CH-1211 Gen\`{e}ve, Switzerland}
\date{\today}
\maketitle

\begin{abstract}
Longitudinal and Hall voltages are measured in a clean
 twinned $YBa_{2}Cu_{3}O_{7-\delta }$ single crystal 
in the liquid and solid vortex phases.
 For magnetic
fields tilted away from the c-axis more than about $2{{}^{\circ }}$, 
a scaling law $\left| \rho _{xy}\right|=A\rho _{xx}^{\beta }$
 with $\beta \approx 1.4$ is observed, which is unaffected by the vortex-lattice
melting transition. The vortex-solid Hall conductivity
 is non-linear and diverges to negative values at low temperature. 
When the magnetic field is aligned to the c-axis,
 the twin-boundary correlated disorder
 modifies the scaling law, 
and $\beta \approx 2$.  The scaling law is unaffected by the Bose-glass
 transition. We discuss the scaling behaviour in terms of the
 dimension-dependent theory for
 percolation in metallic conductors.

\end{abstract}
\vspace{-5pt}
\vspace{-25pt}
\vskip2pc]

A current flowing in a conductor exposed to a magnetic field gives rise to a
Hall voltage. The Hall effect has been a powerful probe of the mechanisms of
charge transport in metals and semiconductors. Similarly, a Hall voltage is
observed in superconductors in high magnetic fields and carrying large
electric currents. The Hall effect in this system is an intriguing
phenomenon which has triggered a very large experimental and theoretical
literature. Remarkable experimental facts include the ''Hall anomaly'',
i.e., the Hall effect sign reversal in the superconducting vortex state with
respect to the normal state, as observed in various high- and
low-temperature type-II superconductors\cite{HagenPRB93Review}, and the
''scaling law'', i.e., the power-law dependence of the Hall resistivity with
respect to the longitudinal resistivity\cite{LuPRL92}.

Many theoretical explanations have been proposed, most of them addressing
the Hall anomaly which is belived to be a fundamental problem of vortex
dynamics. These theories are developed either in terms of microscopic
electronic processes\cite{HirschPRB91}\cite{Feifel'manJETP95}\cite
{KhomskiiPRL95}\cite{KopninPRB95}\cite{WuytsMagneticScatt96}, or including
pinning\cite{WangPRL94}, vortex-vortex interactions\cite{AoVarious}\cite
{JensenEPL92}, time-dependent Ginzburg-Landau theories\cite{DorseyPRB92},
phenomenological models\cite{VinokurPRL93Hall}\cite{DorseyPRL92}\cite
{IkedaJP96}, or other ideas\cite{BrandtREV}. The most frequently adopted
approach is microscopic: it ascribes the Hall effect in the vortex state to
hydrodynamic and vortex-core forces which determine the single vortex
trajectory (e.g. in ref.\cite{Feifel'manJETP95}). In the scenario the Hall
sign reversal results from microscopic details of the Fermi surface. The
situation remains, however, debated and a consensus is not achieved on
fundamental points like the transverse force on a vortex moving in a
superfluid\cite{ThoulessPRL96}\cite{HallComm}, or on experimental problems
like the doping dependence\cite{DopingProb}.

Recently the vortex-lattice melting transition has been shown to influence
the Hall conductivity\cite{D'AnnaPRL98}, which has rised questions to which
extent the microscopic approach of the Hall behavior in the vortex state is
legitimate\cite{AoComment}. We report here new measurements intended to
study the scaling law as the system crosses the vortex-lattice melting
transition, or the Bose-glass transition when twin-boundary correlated
disorder is relevant. Testing the scaling law in different vortex phases
provides insights into the origin of the Hall effect and the mechanisms of
magnetic flux transport in type-II superconductors.

The experiments are performed in a very clean twinned $YBa_{2}Cu_{3}O_{7-%
\delta }$ (YBCO) single crystal in which the characteristic features
associated to the vortex phase transitions are observed. The micro-twinned
crystal has dimensions $0.9\times 0.4mm^{2}$ in the a-b plane, and thickness 
$24\mu m$ in the c-direction. The major twin family is at $45{{}^{\circ }}$
from the long edge of the sample. Some untwinned domains and some twins at $%
90{\ {}^{\circ }}$ from the dominant family are also present. The sample
displays a sharp resistive transition at about $T_{c}=93.5K$. The
longitudinal resistivity $\rho _{xx}$ and Hall resistivity $\rho _{xy}$ are
measured simultaneously by injecting an ac current ($30Hz$), sometimes on
the top of a dc current, along the longest dimension of the crystal, and by
measuring the in-phase voltages parallel and perpendicular to the current.
The experimental method is presented in detail in a previous work on the
same sample\cite{D'AnnaPRL98}. The Hall conductivity is obtained by $\sigma
_{xy}=\rho _{xy}/\left( \rho _{xx}^{2}+\rho _{xy}^{2}\right) $, and the Hall
angle $\theta _{H}$ by $\tan \theta _{H}=\rho _{xy}/\rho _{xx}$.

We begin by discussing the angular dependence and the effect of
twin-boundaries in our sample. The Bose-glass theory\cite{NelsonPRL92PRB93}
predicts that for magnetic fields well aligned to the twin-boundaries the
vortex-solid phase is a smecticlike phase and the transition to the
vortex-liquid is a Bose-glass transition. When the field is tilted away from
the twin-boundaries the vortex-solid phase is a Bragg-glass\cite{Bragg} and
the transition to the vortex-liquid is a vortex-lattice melting transition.

We found experimentally evidence for this angular behavior. Figure 1 shows
the longitudinal resistivity $\rho _{xx}$ measured at $6T$ for zero dc
current and a small ac current of $j_{ac}=1A/cm^{2}$, and for different
angles $\alpha $ between the applied magnetic field and the c-axis, as a
function of the temperature. One can clearly see the effect of
twin-boundaries below $T_{TB}$ . The twin-boundary pinning reduces the
longitudinal resistivity, as expected for correlated disorder\cite
{NelsonPRL92PRB93}. The inset of Fig. 1 shows the onset temperature $%
T_{onset}$ of $\rho _{xx}$ measured with a criterion of $0.1\mu \Omega cm$,
as a function of the angle. For decreasing, large angles $T_{onset}\left(
=T_{m}\right) $ decreases according to a usual anisotropy law\cite
{BlatterREV}. For about $\alpha <2{^{\circ }}$, the onset temperature $%
T_{onset}\left( =T_{BG}\right) $ increases and reaches a maximum at $\alpha
=0{{}^{\circ }}.$ This kind of behavior has been associated\cite
{GrigeraPRL98} to the change in the nature of the transition according to
the Bose-glass theory and the crossover angle is about $2{{}^{\circ }}$ for
twinned YBCO crystals similar to the one we use. Therefore, the onset in
resistivity in Fig. 1 can be associated to the vortex-lattice melting
transition\cite{SafarPRL92}\cite{order} for $\alpha >$ $2{{}^{\circ }}$,
that we denote by $T_{m}$, and to the Bose-glass transition\cite{notaBG} for 
$\alpha <$ $2{{}^{\circ }}$, that we denote by $T_{BG}$.

When the vortex-lattice response is probed by superimposing the ac current
on top of a large dc current, a longitudinal resistivity different from zero
is observed below $T_{m}$ or $T_{BG}$, as the vortex-solid is moving under
the effect of the large Lorentz force. We then also detect a Hall voltage,
and obtain the Hall resistivity and the Hall conductivity in the
vortex-solid phase\cite{D'AnnaPRL98}. The inset of Fig. 2 shows the Hall
conductivity $\sigma _{xy}$ as a function of the temperature at $2T$ and at $%
\alpha =7{^{\circ }}$ and $\alpha =0{^{\circ }}$, measured with large dc and
ac current densities ($j_{dc}=150A/cm^{2},$ $j_{ac}=50A/cm^{2}$) so that the
Hall signal is detected deep inside the vortex-solid. In the inset the small
difference in temperature between the vortex-lattice melting transition at $%
T_{m}$ for $\alpha =7{{}^{\circ }}$ and the Bose-glass transition at $T_{BG}$
for $\alpha =0{{}^{\circ }}$ is not visible. By reducing the temperature
from the normal state the Hall conductivity $\sigma _{xy}$ becomes negative
below $T_{c}$. In the vortex-liquid phase the Hall conductivities at $\alpha
=0{{}^{\circ }}$ and $\alpha =7{{}^{\circ }}$ coincide down to about $%
T_{TB}. $ Below roughly $T_{TB}$ and for $\alpha =0{^{\circ }}$ we observe
an approximately constant Hall conductivity until the large scattering of
the data begins. For the angle tilted away from the c-axis, $\alpha =7{%
^{\circ },}$ the Hall conductivity decreases smoothly until the
vortex-lattice melting transition. Below $T_{m}$ the Hall conductivity
deviates from its behavior in the vortex-liquid phase and goes rapidly
towards large negative values (see also the current dependence in Fig. 3
below). The Hall angle, not shown in Fig. 2, tends to small values.

We investigate now the scaling behavior between the Hall resistivity and the
longitudinal resistivity, that is the existence of a scaling law $\rho
_{xy}\propto \rho _{xx}^{\beta }.$ The main panel of Fig. 2 shows the
log-log plot of $\left| \rho _{xy}\right| $ {\it vs} $\rho _{xx}$ for $%
\alpha =7{{}^{\circ }}$ and $\alpha =0{^{\circ }}$ at $2T$ and large dc and
ac current densities. The position of the vortex-lattice melting and
Bose-glass temperatures are indicated. The fit to a power-law dependence of
a form $\left| \rho _{xy}\right| =A\rho _{xx}^{\beta }$, gives for $\alpha =0%
{^{\circ }}$ the values $A\approx 0.005$ and $\beta \approx 2.0,$ and for $%
\alpha =7{^{\circ }}$ it gives $A\approx 0.02$ and $\beta \approx 1.4$, as
shown by the two straight dotted lines. A separate fit to the solid and
liquid part gives the same result within the experimental error (we also
obtain $\beta \approx 1.4$ in the whole range $3{{}^{\circ }}$ to $7{%
{}^{\circ }}$). {\it There is no change of the }$\left| \rho _{xy}\right| $ 
{\it vs} $\rho _{xx}${\it \ dependence at the vortex-lattice melting
transition or at the Bose-glass transition}, suggesting that such a scaling
law is effectively insensitive to the specific vortex phase.

The Hall effect current dependence is shown in Fig. 3 for $\alpha =3{%
{}^{\circ }}$. The inset of Fig. 3 shows the Hall conductivity $\sigma _{xy}$
as a function of the magnetic field at $89K$ and different dc currents. At $%
\alpha =3{{}^{\circ }}$ the vortex phase transition is the vortex-lattice
melting at $B_{m}$. The curves have larger noise over signal ratio than
above. Nevertheless the current dependence is clearly observable in the
diverging $\sigma _{xy}$. {\it Below the melting field }$B_{m}${\it \ the
Hall conductivity }$\sigma _{xy}${\it \ decreases faster, the smaller the dc
current}. Above $B_{m}$ in the vortex-liquid the Hall conductivity is
linear. The current dependence of the scaling law is investigated in the
main panel of Fig. 3, which shows a log-log plot of $\left| \rho
_{xy}\right| $ {\it vs} $\rho _{xx}$, constructed from measurements as a
function of the magnetic fields at a constant temperature of $89K$ and
different current densities. The fit to a power-law dependence of a form $%
\left| \rho _{xy}\right| =A\rho _{xx}^{\beta }$, with $A$ and $\beta $ free
parameters as above, gives the average values $A\approx 0.012$ and $\beta
\approx 1.4.$ {\it There is no change of the scaling law with the current
density}, neither in liquid nor in the solid vortex phases.

The data presented here prove that the general trend of the Hall
conductivity is indeed captured by a very robust scaling law $\left| \rho
_{xy}\right| =A\rho _{xx}^{\beta }$. The scaling law implies $\tan \theta
_{H}\propto \rho _{xx}^{\beta -1}$ and $\sigma _{xy}\propto \rho
_{xx}^{\beta -2}$. Consistently, with an exponent less than two at $\alpha >2%
{{}^{\circ }}$, as the longitudinal resistivity tends to zero, the Hall
conductivity diverges, and the Hall angle is small. The strong non-linear
dependence of the longitudinal resistivity in the vortex-solid phase is
reflected in the Hall conductivity, which below the melting transition
diverges faster, the smaller the current density. For $\alpha =0{{}^{\circ }}
$ and consistently with an exponent $\beta \approx 2$ the Hall conductivity
seems to be a constant below about $T_{TB}$, the temperature of
twin-boundary pinning onset.

A Hall resistivity which vanishes as a power of the longitudinal
resistivity, $\rho _{xy}\propto \rho _{xx}^{\beta },$ has been observed by
various authors\cite{LuPRL92}\cite{BudaniVario}\cite{WoltFilms}, and
predicted in different theoretical contexts. Dorsey et al.\cite{DorseyPRL92}
developed a scaling theory near the vortex-glass transition with a power $%
\beta <2$ for the three-dimensional regime. In the model the exponent $\beta 
$ is universal, but the sign of the Hall effect is material specific and
possibly related to microscopic pairing processes. Vinokur et al.\cite
{VinokurPRL93Hall} proposed that the scaling law with $\beta =2$ is a
general feature of any vortex state with disorder-dominated dynamics,
without the need to invoke the vortex-glass scaling. In this model the Hall
conductivity is independent of disorder and is directly linked to the
microscopic processes determining the single vortex equation of motion. Wang
et al.\cite{WangPRL94} have proposed that the pinning affects the single
vortex trajectory via the backflow current inside the normal vortex-core.
This modifies the exponent in such a way that $\beta =1.5$ for strong
pinning and $\beta =2$ for weak pinning, and the Hall sign reversal is a
pinning effect. Ao\cite{AoVarious} derived the scaling law in the context of
a vortex many-body linear theory, where the Hall voltage results from the
motion of vortex-lattice defects (vacancies), with $\beta =2$ for pinning
induced vacancies and $\beta =1$ for thermally (fluctuation) induced
vacancies.

A complete explanation of the puzzling vortex-Hall behavior is not yet
achieved. A consistent theory should explain, in addition to the Hall effect
sign reversal, the robustness of the scaling $\rho _{xy}\propto \rho
_{xx}^{\beta }$ reported here. We have found that $\sigma _{xy}$ becomes
current dependent in the vortex-solid phase. This contradicts the idea that
the Hall conductivity is independent of disorder\cite{VinokurPRL93Hall}. 
{\it Pinning effects have to be considered}. However, even if the
temperature and field dependence of $\sigma _{xy}$ change at vortex phase
transitions (either vortex-lattice melting or Bose-glass transitions), the
scaling law is found to remain independent of the specific vortex phase.
This suggests that a comprehensive theory for the Hall effect far enough
from the sign change does not require phase-dependent parameters. Provided
it reproduces a general scaling law $\rho _{xy}\propto \rho _{xx}^{\beta }$
and leads to the correct sign of $\sigma _{xy}$, the Hall behavior is then
completely determined by the longitudinal resistivity, which englobes
many-body effects, e.g., collective pinning in the vortex-solid phase.

A significant result of this paper is that the exponent $\beta $ entering
the scaling law is {\it disorder-type dependent}. In particular $\beta 
\approx 2.0$ for correlated planar disorder and $\beta \approx 1.4$ for
uncorrelated point disorder. This suggests an alternative explanation for
the scaling behaviour, as proposed by Geshkenbein\cite{GeshPerco}. If one
views the vortex freezing as an inhomogeneous, non simultaneous process,
with regions where vortices are pinned (thus with vanishing resistivity),
and regions where they can still move (thus inducing a non zero electric
resistivity), the vortex freezing behaviour in superconductors has strong
analogies with the {\it percolation transition} in inhomogeneous conductors.
In the case of a mixed metallic/insulating system, the conductivity is
governed by percolation processes and the longitudinal conducticity is
expressed as $\sigma _{xx}\propto \delta p^{t},$ where $\delta p=p-p_{c}$ is
the difference between the conducting metallic phase density $p$ and the
critical percolation threshold density $p_{c}$. The critical exponent is $t%
\approx 1.3$ in two dimensions, and $t\approx 1.6$ in three dimensions\cite
{PercolationCOnductors}. Similarly, the Hall number $R_{H}$ diverges as $%
R_{H}\propto \delta p^{-g},$ where $g=\nu \left( d-2\right) $,\cite
{PercolationCOnductors} where $d$ is the dimension. In two dimensions $g=0$
and as consequence the Hall conductivity $\sigma _{xy}\approx HR_{H}\sigma
_{xx}^{2}$ is exactly proportional to $\sigma _{xx}^{2}$ (at constant
magnetic field $H$). In three dimensions, $g=\nu \approx 0.9$,\cite
{PercolationCOnductors} such that $\sigma _{xy}\propto \sigma
_{xx}^{2-g/t}\propto \sigma _{xx}^{1.44}.$

This immediatly leads to a percolation model for the vortex scaling
behaviour, provided the vortex conductivity is interpreted as the electric
resistivity of the metallic/insulating system, since a high vortex mobility
means large electric dissipation. With the identification that the
conductivity $\sigma $ in the metallic/insulating system is the resistivity $%
\rho $ in the vortex system, one obtains the scaling law $\rho _{xy}\propto
\rho _{xx}^{2-g/t}$ and $\beta =2-g/t$ for the vortex system. The
percolation model is very appealing since it provides two universal
exponents, i.e., $\beta =2$ and $\beta =1.44$, which correspond to the most
frequently reported experimental estimate. These exponents are determined by
the dimensionality of the vortex system, that is, determined by the
intrinsic anisotropy of the material, or by the vortex localization along
correlated defects, or by the geometry of the samples. For the magnetic
field accurately aligned to the twin-boundaries, which localizes the
vortices along the c-axis, the system is two dimensional and $\beta \approx 2
$. The same exponent is observed in two-dimensional films\cite{WoltFilms}.
When the magnetic field is ''slightly'' tilted away from the twin-boundaries
(2 to 3 degrees are enough), the vortices recover the third degree of
freedom and $\beta \approx 1.4$. This is also likely to happen when splayed
defects are introduced by irradiation, bringing back the exponent form $2$
to $1.5$.\cite{BudaniVario}

The vortex-percolation model is indeed very interesting since it predicts a
scaling law independent of the vortex phase, as observed here, and explains
the robustness of the scaling since it is related only to the dimensionality
of the vortex system.

We are grateful to V. Geshkenbein and G. Blatter for many discussions. This
work is supported by the Swiss National Science Foundation.

\bigskip

\begin{center}
{\bf Figure captions}
\end{center}

FIG. 1. The longitudinal resisitivity $\rho _{xx}$ in a twinned $%
YBa_{2}Cu_{3}O_{7-\delta }$ single crystal at $6T$ as a function of the
temperature, measured at low ac current density, $j_{ac}=1A/cm^{2}$, for
different angles between the field and the c-axis (-0.2${{}^\circ},$ 0${%
{}^\circ},$ 0.2${{}^\circ},$ 0.5${{}^\circ},$ 1${{}^\circ},$ 2${{}^\circ},$ 3%
${{}^\circ},$ 5${{}^\circ},$ 7${{}^\circ},$ 10${{}^\circ},$ 12${{}^\circ},$
20${{}^\circ})$. Inset: the onset temperature $T_{onst}$ as a function of
the angle, at $2T$ and $6T$. For about $\alpha >2{{}^{\circ }}$ the onset
temperature follows the usual anisotropic law of the vortex-lattice melting
temperature. For small angles the onset temperature increases as expected
from the Bose-glass theory (see text for details).

FIG. 2. Log-log plot of $\left| \rho _{xy}\right| $ versus $\rho _{xx}$ at $%
2T$ and $\alpha =7{^{\circ }}$ and $\alpha =0{{}^{\circ }}$. The linear fit
according to a scaling law $\left| \rho _{xy}\right| =A\rho _{xx}^{\beta }$
gives $\beta \approx 1.4$ for $\alpha =7{{}^{\circ }}$, and $\beta \approx
2.0$ for $\alpha =0{{}^{\circ }},$ as indicated by the two straigth dotted
lines. The position of the vortex-lattice melting temperature, $T_{m},$ and
of the Bose-glass temperature, $T_{BG}$, are indicated in the curves, as
well as the temperature of twin-boundary pinning onset, $T_{TB}.$ Notice
that the scaling law is unaffected by crossing the transitions. Inset: the
Hall conductivity, $\sigma _{xy}=\rho _{xy}/\left( \rho _{xx}^{2}+\rho
_{xy}^{2}\right) $, as a function of the temperature at $\alpha =7{^{\circ }}
$ and $\alpha =0{{}^{\circ }.}$ The dotted vertical lines denote the
transitions at $T_{m}$, $T_{BG}$ and $T_{TB}$. The current densities are $%
j_{dc}=150A/cm^{2},$ $j_{ac}=50A/cm^{2}$.

FIG. 3. Log-log plot of $\left| \rho _{xy}\right| $ versus $\rho _{xx}$ for
different current densities at $89K$ and $\alpha =3{^{\circ }}$. For each
curve the current densities are indicated by $(j_{dc},j_{ac})$, both in unit
of $A/cm^{2}$. The position of the vortex-lattice melting field is
indicated. The linear fit according to a scaling law $\left| \rho
_{xy}\right| =A\rho _{xx}^{\beta }$ gives the average values $A\approx 0.012$
and $\beta \approx 1.4$ (see dotted line) and there is no current dependence
of the parameters. Inset: The field dependence of the Hall conductivity $%
\sigma _{xy}$ at $89K$, for various ac and dc current densities. The dotted
vertical line denotes the vortex-lattice melting transition at $B_{m}$.


\begin{references}
\bibitem{notaBG}  Because some parts of our sample seem untwinned, for $%
\alpha =0{{}^{\circ }}$ the transition is possibly not completely determined
by the correlated disorder.
\end{references}
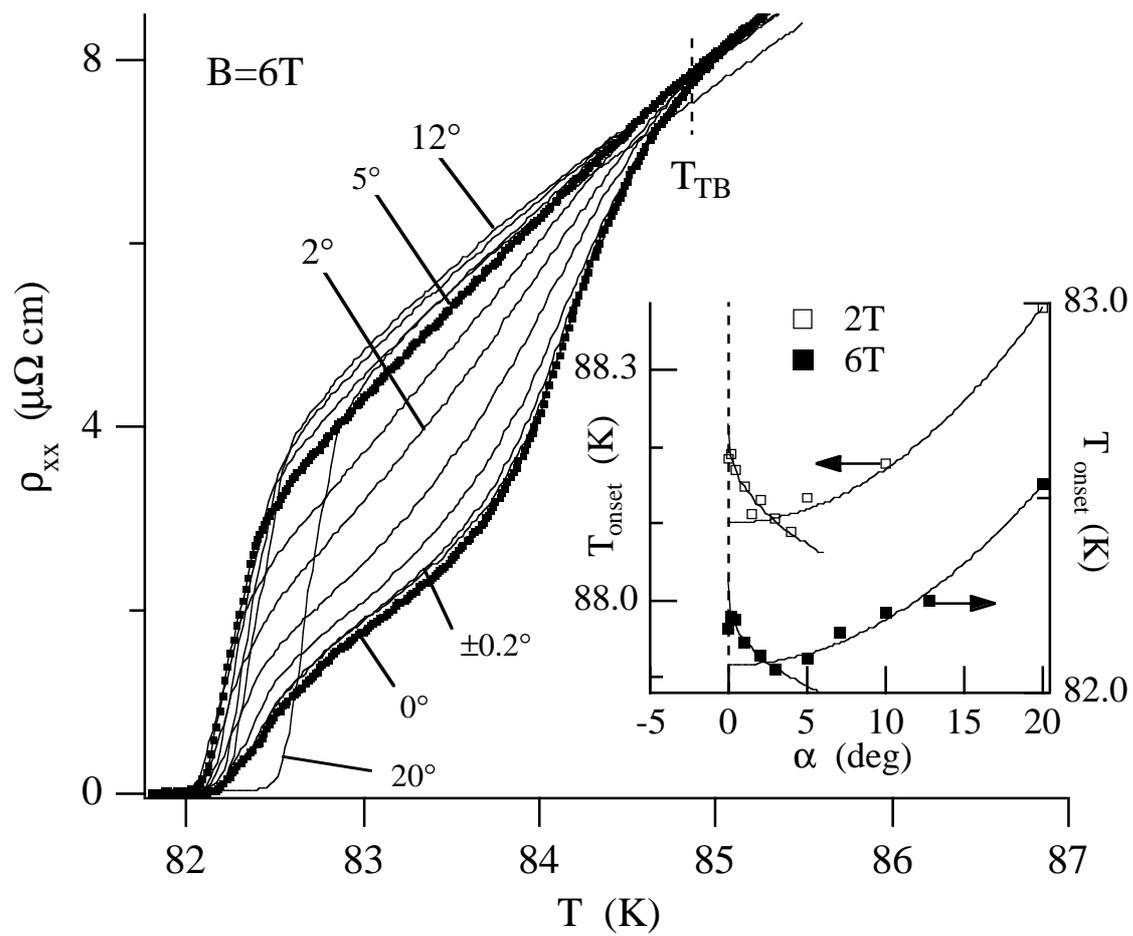
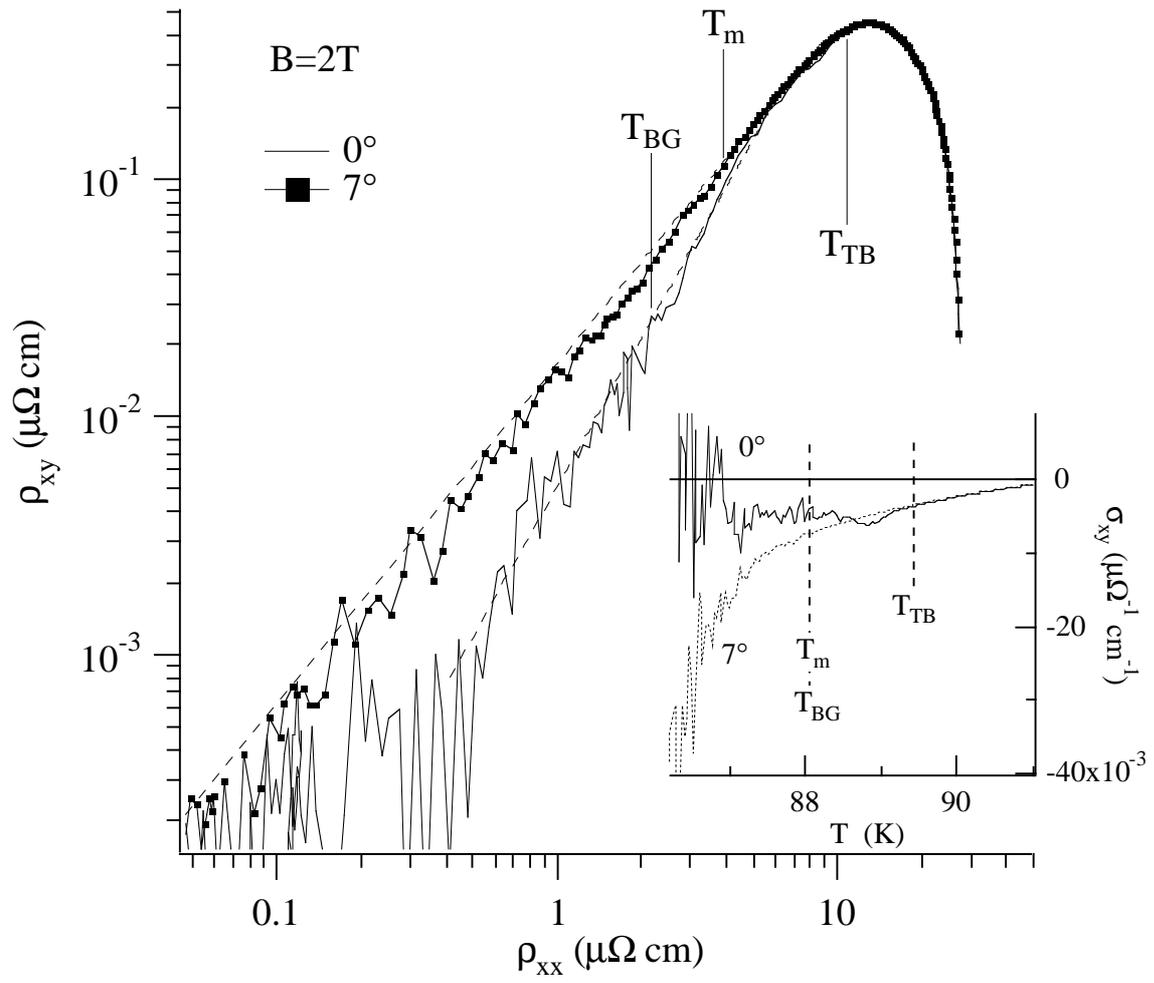
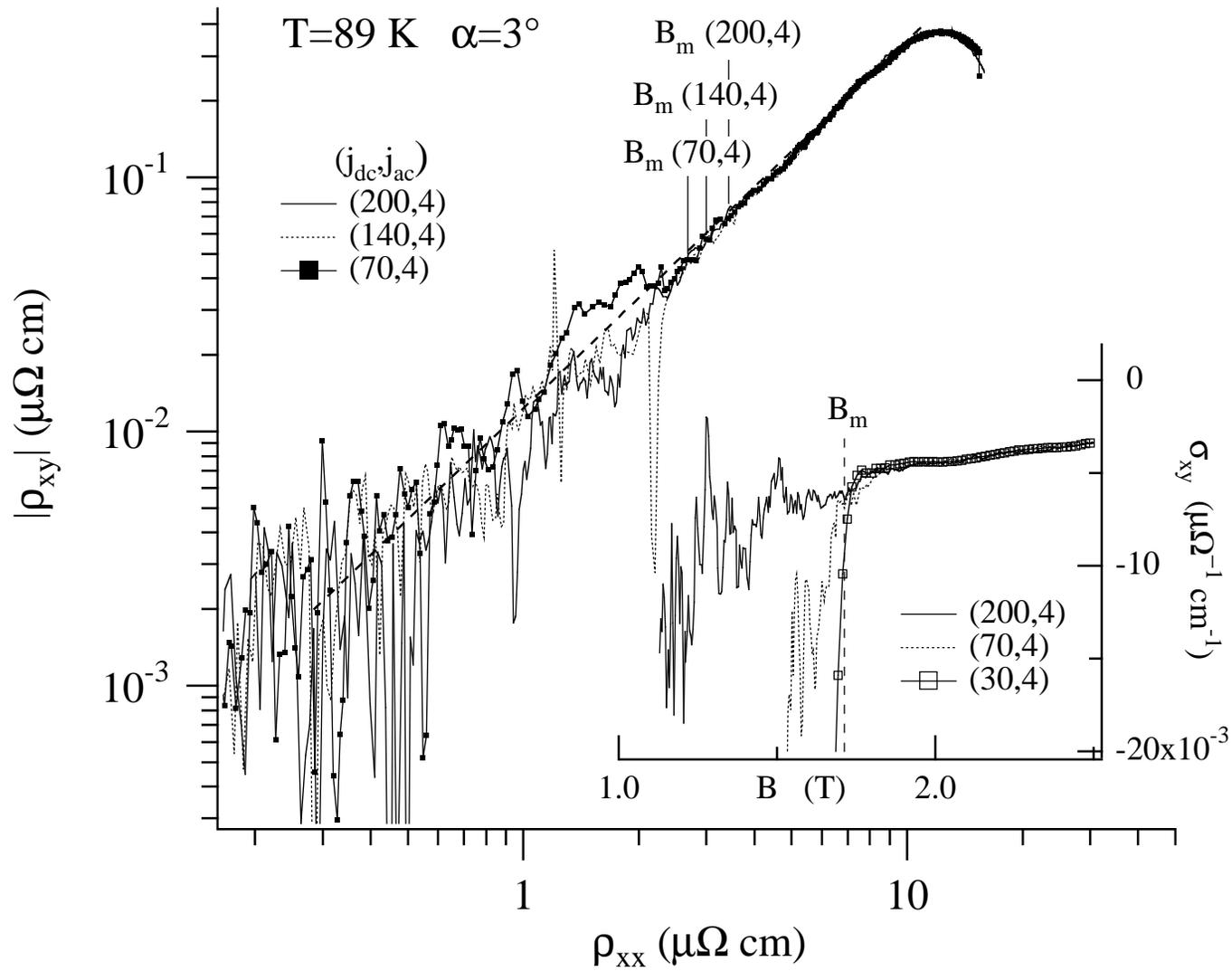
\end{document}